# The dynamic nature of trust: Trust in Human-Robot Interaction revisited

Jimin Rhim[1*]

University of Ottawa, School of Engineering Design and Teaching Innovation, jrhim@uottawa.ca

SONYA S. KWAK

KIST, Center for Intelligent and Interactive Robotics, sonakwak@kist.re.kr

ANGELICA LIM

Simon Fraser University, School of Computing Science, angelica@sfu.ca

JASON MILLAR

University of Ottawa, School of Engineering Design and Teaching Innovation, jmillar@uottawa.ca

The role of robots is expanding from tool to collaborator. Socially assistive robots (SARs) are an example of collaborative robots that assist humans in the real world. As robots enter our social sphere, unforeseen risks occur during human-robot interaction (HRI), as everyday human space is full of uncertainties. Risk introduces an element of trust, so understanding human trust in the robot is imperative to initiate and maintain interactions with robots over time. While many scholars have investigated the issue of human-robot trust, a significant portion of that discussion is rooted in the human-automation interaction literature. As robots are no longer mere instruments, but social agents that co-exist with humans, we need a new lens to investigate the longitudinal dynamic nature of trust in HRI. In this position paper, we contend that focusing on the dynamic nature of trust as a new inquiry will help us better design trustworthy robots.

**CCS CONCEPTS • Human-centered computing →HCI theory, concepts and models**

**Additional Keywords and Phrases:** Socially assistive robots (SARs), human-robot interaction, trust, trustworthy robotics

## 1. INTRODUCTION

Conventional robotic arms were placed in manufacturing settings to conduct repetitive, dull, and dangerous tasks for humans. Typically, only trained users have direct access to interact with such robots due to safety concerns. These days, however, more and more robotic arms are being deployed to interact with laypeople. For instance, barista robots make and serve coffee for customers, and assistive robots interact with patients with disabilities [7]. While these robot arms provide convenience and novel user experiences, and tend to be smaller and less physically dangerous, they can still create unexpected hazards or confusion when interacting with untrained human users—the wide range of movements of robotic arms makes it difficult for humans to fully anticipate their movements. As the aforementioned example indicates, we are witnessing an increase in the application of robots in our social spheres where humans interact with robots in everyday tasks (e.g., mobility [24], adopted in healthcare [22], entertainment [1], and education [4] sectors, and act as social assistance for older adults [27]). The expansion of robot roles from tools to teammates poses novel questions regarding the co-existence of humans and robots. This shift leads to a state of affairs in which robots and humans co-exist, which indicates that a broader range of lay human users will face various types and levels of risk or hazard during HRI. Risk and uncertainty inevitably involve trust [15], so understanding the longitudinal aspects of human trust in robot partners is imperative to initiate and maintain relationships with robots over time.

Many scholars have highlighted that a comprehensive conceptualization of trust is essential when designing robots that interact socially with humans because trust is integral for a user's acceptance and inclusion of the robot into their social sphere [23]. That is, a user is unlikely to use a robot if they believe that the robot is untrustworthy.

---

While trust can induce cooperation between humans and robots, forming well-placed trust is extremely difficult. Further, misaligned trust toward a robot leads to the misuse or disuse of a robot. For instance, people tend to misuse a robot when they over-trust it or disuse a robot when they under-trust it [12]. Several studies have shown the caveats of misplacing trust in robots: one study showed that participants mindlessly followed instructions from a robot during potentially dangerous situations even if the robot made risky and unsafe suggestions [25], which had the potential to endanger the user. While the consequences of disusing a robot tend to have less direct harm than the misuse of a robot, there remain negative impacts of robot disuse, including that users may not take advantage of the potential benefits a robot could provide [32]. The above cases specify that successful human-robot trust formation aims to form a well-calibrated trust during HRI.

A broad spectrum of literature has developed with increasing emphasis on the importance of human-robot trust. Topics on the human-robot trust include trust measurement [21], trust repair strategies [3,9], and trust modelling [29]. However, the foundation of empirical and theoretical literature on trust in robots is often centred around the automated system [12,19]. While the insights gained from studies that regard trust in automation provide profound knowledge for understanding trust in robots, the trust characteristics of human interactions with SARs have different implications than considerations regarding human interactions with automated machines. A robot is distinguished from an automated machine or computer interface by the embodiment and corporeality of robots with multi-modal interaction which can provide richer social interactions with humans [13]. Such dissimilarities highlight the importance of grounding human-robot trust beyond literature regarding human-automation trust. As the roles and expectations of robots change from tools to collaborators, we need to revisit the notion of trust which is more appropriate for the context of HRI. Next, we provide an outline of the gaps in the literature to identify the absence of an adequate conceptualization of trust in current deliberations in HRI. We then identify different layers of the dynamic nature of trust over time (i.e. the longitudinal aspects of trust and trustworthiness) during HRI to open up discussions for the future research directions.

## 2. GAPS IN CURRENT HUMAN-ROBOT TRUST RESEARCH

What does it mean for humans to trust a robot? While extensive literature discusses various aspects that impact human-robot trust over the past two decades, there is no clear agreement on the definition of trust in HRI. Further, there is a discrepancy between the current public perception of trust toward robots and what is articulated in research. While many studies treat trust towards robots as something that has already been established, many contemporary reports indicate that the public is reluctant to trust robots, leading to some uncertainty whether people trust robots. In response, we posit that the ill-posed nature of the current discussion of human-robot trust in research hinders the development of well-placed trust between humans and robots. In this section, we outline several factors that may impede concretizing the notion of trust during HRI to shed light on how we could approach to overcome these challenges.

### 2.1. A LACK OF CONSENSUS IN THE DEFINITION OF TRUST IN HRI

According to Salem et al., 2015 [26], the concept of trust is still an ongoing exploration in HRI due to the sheer complexity of the concept itself. One challenge that makes the concept of trust elusive is that there are multiple synonyms for trust. Consider the interchangeability of the terms e.g., reliability, faith, confidence, belief, vulnerability, certainty, and credit with trust. Another challenge is the difficulty of coining the term trust in the specific context of HRI. Yet, another challenge is the difficulty of coining the term trust in the specific context of HRI. It is widely accepted that different disciplines define and analyze the concept of trust differently [5]. In parallel, the focal point of the trust discussions in human-machine interaction differs depending on the medium and the context of its deployment. For instance, the literature on trust in human-automation emphasizes the reliability, robustness, predictability, and safety of automated systems as core factors that shape human trust [12]. This outlook portrays the role of the humans as operators who expect machines to function predictably. On the other hand, the discussion on trust for robots that interact socially with humans includes considerations of user characteristics (e.g., social factors, user propensity) [11] on top of the reliability of the robot's functionality. The expansion of robot roles and user expectations in social contexts implies that human-robot trust discussions should avoid monolithic perspectives and should rather reflect the multifaceted nature of trust in HRI.

### 2.2. GAP BETWEEN ROBOT-CENTRIC HRI VERSUS HUMAN-CENTRIC HRI

HRI's treatment of trust appears to be divided into two main categories: human-centric and robot-centric perspectives [14,16]. Human-centered HRI research investigates themes such as design or usability of robots often through user studies, whereas robot-centered HRI research investigates algorithms and engineering innovations that improve the overall performance of the robot [14]. In terms of trust studies, human-centric HRI explores the factors that affect human users' perception of trust (e.g., demographics, personality traits, attitudes toward robots, and propensity to trust). One such study investigated the impact of erroneous robot behaviour on human subjective perception and acceptance of the robot's trustworthiness [26]. An example of a robot-centric HRI trust study includes one that developed an Online Probabilistic Trust Inference Model (OPTIMo)—a widely adopted computational model that estimates near real-time human trust toward a robot by observing human behaviours [31]. However, as the examples indicate, despite the common intention to shape human-robot trust, each research approach differs depending on the focus of perspective. Accordingly, we need to have a context-dependent understanding of trust to provide appropriate discussions for the trust discussion at hand.

### 2.3. GAP BETWEEN VIEWING HUMAN-ROBOT TRUST RESEARCH AS A VARIABLE VERSUS PROCESS

Trust in HRI is a multifaceted concept with many layers and a dynamic process that fluctuates over time. The temporal trust trajectory includes the following phases: development, dissolution, and restoration [18]. However, many empirical studies in human-robot trust treat trust as a time-independent variable, often measuring the level of trust through surveys and experiments [6]. Studies that treat trust as an independent variable tend to focus on the benefit of trust [17], such as how trust facilitates cooperation with humans or reduces uncertainties during HRI [30]. Studies that view trust as a dependent variable focus on factors that directly impact trust [17], such as users' attitudes towards robots, operator's performance, and failure rates of robots [10]. Conventional study methods that treat trust during HRI as variables may not fully capture the characteristic of trust that encapsulates temporality and dynamic nature. Nonetheless, if we envision longitudinal interactions with the robot, trust should be explained as a process rather than a variable.

### 3. THE DYNAMIC NATURE OF TRUST DURING HRI

The traditional discussion of trust during HRI needs to fully convey the dynamic nature of trust between the human who trusts and the robot that is trusted. As elaborated in a previous section, the meaning of human-robot trust does not have a clearly established definition, which implies that a new conception needs to be formed through inferences [28]. This section provides an account of how different layers of the dynamic nature of trust should be considered when establishing a conception of human-robot trust by drawing discussions from multidisciplinary research fields.

First, the dynamic act of trusting relies upon social interaction. As previously posited, most of the trust discussions are based on the discussion of human-automation interaction, where the focus of trust formation is based on the robustness of the robot's performance. This view treats trust as instrumental or consequential. However, as robots become collaborators that conduct various social tasks alongside humans, it is likely that people will not only expect functional success from robots but also expect robots to fulfill social expectations (e.g., how empathetic the robot is, how well the robot respects social norms). As such, it is imperative for HRI to include trust as a social and emotional act that considers the relationship between humans and robots. A human-human interaction study that regards the social dynamic of trust illustrates this distinction: "people's decision to trust is best predicted by the emotions they attach to the action itself rather than by emotions they attach to possible outcomes" (p. 692) [8]. An HRI study also found how people perceive a robot to be more significant for trust formation than the robot's performance alone [26]. To that end, human-robot trust theories should also include the dynamic social aspects of human nature, including factors such as context- and time-dependent social norms, relationship status, and emotions.

Second, HRI should address the temporal dynamic of trust. Temporal dynamics refer to the fluctuation of human trust towards a robot over time, and the factors that impact trust differs over time. Stage models of trust [18] delineate how different factors impact trust development or deteriorate in different stages [2,20]. Calculus-based trust is most prevalent in the initial trust formation stage [20]. The most critical facet at this stage is that the trustor (the human) can be assured that the trustee (the robot) will act according to his or her expectations; thus, reliability and dependability are integral. The latter trust formation stage relies heavily on knowledge-based trust: trust is based on accumulated knowledge of the trustee's ability over repeated interactions. The third stage involves identification-based trust, where the trustor expects the trustee's values or interests to align with their own. This means that calculus-based trust is not as important at the latter stage of trust, but rather the perception of the trustworthiness of a trustee is more nuanced. As this stage model indicates, it is important to consider that human-robot trust may fluctuate over repeated interactions, and different factors will have different implications depending on the stage of trust formation.

### 4. CONCLUSION

The current discussion of trust in robotics tends to conceive trust as a unitary, fully established, and stable value. However, a careful exploration of the concept thus reveals that trust is a dynamic concept in a state of flux. This position paper highlights the versatile and dynamic nature of trust that should be considered to progress realistic discussion of trust during HRI. In summary, it is important to consider that trust is not a stable state during HRI. Trust is dynamic in both temporal and social aspects. One significant hindrance to conceptualizing human-robot trust as dynamic in nature is that most empirical studies conducted in HRI to understand trust are based on one-shot study designs. Consequently, existing studies do not allow for the examination of growth or decrease of trust over time. These studies inevitably tend towards highlighting calculus-based trust. This approach naturally eliminates opportunities to examine the social dynamics of trust during HRI. While it is essential for robots to ensure their behaviours are trustworthy in order to initiate interaction, consideration of only the early stages of trust may not convey the full process of trust development during HRI. As such, we propose that when designing trustworthy robots that depend on trust formation, the consideration of a variety of different factors is crucial. Consequently, longitudinal studies that model trust trajectories are necessary for understanding the dynamic nature of trust comprehensively.

**ACKNOWLEDGMENTS**

We would like to thank Dr. Jung-Mi Park for sharing some examples of robots that are adopted in the wild.

**REFERENCES**

[1] Iina Aaltonen, Anne Arvola, Päivi Heikkilä, and Hanna Lammi. 2017. Hello Pepper, may I tickle you?: 2017 ACM/IEEE International Conference on Human-Robot Interaction. *HRI '17*: 53–54.

[2] M. Audrey Korsgaard, Jason Kautz, Paul Bliese, Katarzyna Samson, and Patrycjusz Kostyszyn. 2018. Conceptualising time as a level of analysis: New directions in the analysis of trust dynamics. *Journal of Trust Research* 8, 2: 142–165.

[3] Anthony Baker, Elizabeth Phillips, Daniel Ullman, and Joseph Keebler. 2018. Toward an Understanding of Trust Repair in Human-Robot Interaction: Current Research and Future Directions. *ACM Transactions on Interactive Intelligent Systems* 8: 1–30.

[4] Paul Baxter, Emily Ashurst, Robin Read, James Kennedy, and Tony Belpaeme. 2017. Robot education peers in a situated primary school study: Personalisation promotes child learning. *PLOS ONE* 12, 5: e0178126.

[5] Kirsimarja Blomqvist. 1997. The many faces of trust. *Scandinavian Journal of Management* 13, 3: 271–286.

[6] Rik van den Brule, Ron Dotsch, Gijsbert Bijlstra, Daniel H. J. Wigboldus, and Pim Haselager. 2014. Do Robot Performance and Behavioral Style affect Human Trust? *International Journal of Social Robotics* 6, 4: 519–531.

[7] Alexandre Campeau-Lecours, Hugo Lamontagne, Simon Latour, et al. 2019. Kinova Modular Robot Arms for Service Robotics Applications. *Rapid Automation: Concepts, Methodologies, Tools, and Applications*, 693–719. Retrieved February 14, 2023 from https://www.igi-global.com/chapter/kinova-modular-robot-arms-for-service-robotics-applications/www.igi-global.com/chapter/kinova-modular-robot-arms-for-service-robotics-applications/222454.

[8] David Dunning, Detlef Fetchenhauer, and Thomas M. Schlösser. 2012. Trust as a social and emotional act: Noneconomic considerations in trust behavior. *Journal of Economic Psychology* 33, 3: 686–694.

[9] Connor Esterwood and Lionel P. Robert. 2021. Do You Still Trust Me? Human-Robot Trust Repair Strategies. *2021 30th IEEE International Conference on Robot & Human Interactive Communication (RO-MAN)*, 183–188.

[10] P. A. Hancock, Theresa T. Kessler, Alexandra D. Kaplan, John C. Brill, and James L. Szalma. 2021. Evolving Trust in Robots: Specification Through Sequential and Comparative Meta-Analyses. *Human Factors* 63, 7: 1196–1229.

[11] Peter A. Hancock, Deborah R. Billings, Kristin E. Schaefer, Jessie Y. C. Chen, Ewart J. de Visser, and Raja Parasuraman. 2011. A Meta-Analysis of Factors Affecting Trust in Human-Robot Interaction. *Human Factors* 53, 5: 517–527.

[12] Kevin Anthony Hoff and Masooda Bashir. 2015. Trust in Automation: Integrating Empirical Evidence on Factors That Influence Trust. *Human Factors* 57, 3: 407–434.

[13] Laura Hoffmann, Nikolai Bock, and Astrid M. Rosenthal v.d. Pütten. 2018. The Peculiarities of Robot Embodiment (EmCorp-Scale): Development, Validation and Initial Test of the Embodiment and Corporeality of Artificial Agents Scale. *Proceedings of the 2018 ACM/IEEE International Conference on Human-Robot Interaction*, Association for Computing Machinery, 370–378.

[14] Roohollah Jahanmahin, Sara Masoud, Jeremy Rickli, and Ana Djuric. 2022. Human-robot interactions in manufacturing: A survey of human behavior modeling. *Robotics and Computer-Integrated Manufacturing* 78: 102404.

[15] Audun Jøsang and Stéphane Lo Presti. 2004. Analysing the Relationship between Risk and Trust. In C. Jensen, S. Poslad, and T. Dimitrakos, eds., *Trust Management*. Springer Berlin Heidelberg, Berlin, Heidelberg, 135–145.

[16] Zahra Rezaei Khavas, S. Reza Ahmadzadeh, and Paul Robinette. 2020. Modeling Trust in Human-Robot Interaction: A Survey. *Social Robotics*, Springer International Publishing, 529–541.

[17] Dmitry Khodyakov. 2007. Trust as a Process: A Three-Dimensional Approach. *Sociology* 41, 1: 115–132.

[18] Peter H. Kim, Kurt T. Dirks, and Cecily D. Cooper. 2009. The Repair of Trust: A Dynamic Bilateral Perspective and Multilevel Conceptualization. *Academy of Management Review* 34, 3: 401–422.

[19] John D. Lee and Katrina A. See. 2004. Trust in Automation: Designing for Appropriate Reliance. *Human Factors* 46, 1: 50–80.

[20] Roy J. Lewicki, Edward C. Tomlinson, and Nicole Gillespie. 2006. Models of Interpersonal Trust Development: Theoretical Approaches, Empirical Evidence, and Future Directions. *Journal of Management* 32, 6: 991–1022.

[21] Bertram F. Malle and Daniel Ullman. 2021. Chapter 1 - A multidimensional conception and measure of human-robot trust. In C.S. Nam and J.B. Lyons, eds., *Trust in Human-Robot Interaction*. Academic Press, 3–25.

[22] Jordan A. Mann, Bruce A. MacDonald, I. -Han Kuo, Xingyan Li, and Elizabeth Broadbent. 2015. People respond better to robots than computer tablets delivering healthcare instructions. *Computers in Human Behavior* 43: 112–117.

[23] Raja Parasuraman and Victor Riley. 1997. Humans and Automation: Use, Misuse, Disuse, Abuse. *Human Factors* 39, 2: 230–253.

[24] Jürgen Pripfl, Tobias Körtner, Daliah Batko-Klein, et al. 2016. Results of a real world trial with a mobile social service robot for older adults. *2016 11th ACM/IEEE International Conference on Human-Robot Interaction (HRI)*, 497–498.

[25] Paul Robinette, Alan R. Wagner, and Ayanna M. Howard. 2016. Investigating human-robot trust in emergency scenarios: methodological lessons learned. In *Robust intelligence and trust in autonomous systems*. Springer, 143–166.

[26] Maha Salem, Gabriella Lakatos, Farshid Amirabdollahian, and Kerstin Dautenhahn. 2015. Would You Trust a (Faulty) Robot? Effects of Error, Task Type and Personality on Human-Robot Cooperation and Trust. *2015 10th ACM/IEEE International Conference on Human-Robot Interaction (HRI)*, 1–8.

[27] Cory-Ann Smarr, Tracy L. Mitzner, Jenay M. Beer, et al. 2014. Domestic Robots for Older Adults: Attitudes, Preferences, and Potential. *International Journal of Social Robotics* 6, 2: 229–247.

[28] Herman Veluwenkamp and Jeroen van den Hoven. 2023. Design for values and conceptual engineering. *Ethics and Information Technology* 25, 1: 2.

[29] Samuele Vinanzi, Massimiliano Patacchiola, Antonio Chella, and Angelo Cangelosi. 2019. Would a robot trust you? Developmental robotics model of trust and theory of mind. *Philosophical Transactions of The Royal Society B Biological Sciences* 374.

[30] Jane Wu, Erin Paeng, Kari Linder, Piercarlo Valdesolo, and James Boerkoel Jr. 2016. *Trust and Cooperation in Human-Robot Decision Making*. .

[31] Anqi Xu and Gregory Dudek. 2015. OPTIMo: Online Probabilistic Trust Inference Model for Asymmetric Human-Robot Collaborations. *Proceedings of the Tenth Annual ACM/IEEE International Conference on Human-Robot Interaction*, Association for Computing Machinery, 221–228.

[32] Jin Xu and Ayanna Howard. 2018. The Impact of First Impressions on Human- Robot Trust During Problem-Solving Scenarios. *2018 27th IEEE International Symposium on Robot and Human Interactive Communication (RO-MAN)*, 435–441.